\documentclass[
reprint,
a4paper,
secnumarabic,
showpacs,showkeys,preprintnumbers,
nofootinbib,
 amsmath,amssymb,
 aps,
]{revtex4-1}

\usepackage{graphicx}
\usepackage{bm}

\newcommand{\rmd}{\mathrm{d}}
\newcommand{\rmi}{\mathrm{i}}

\begin{document}

\title{Set Theory as the Unified Scheme for Physics}

\author{A.V.~Novikov-Borodin}
\affiliation{Institute for Nuclear Research of RAS, 60-th October Anniversary prospect~7a, 117312 Moscow, Russia}
\begin{abstract} 
The process of cognition is analysed to adjust the set theory to physical description. Postulates and basic definitions are revised. The specific sets of predicates, called presets, corresponding to the physical objects identified by an observer during cognition are introduced. Unlike sets, the presets are free of logical or set-theoretical paradoxes and may be consistently used in physical description. Schemes of cognition based on presets are considered. Being different logical systems, the relativistic and quantum theories, observations in modern cosmology cannot be consistently considered in one `unified physical theory', but they are in frames of introduced schemes of cognition. 
\end{abstract}
  
\maketitle
%
\section*{\bf Introduction}

The set theory is a foundation from which virtually all of mathematics can be derived. Since mathematics is a language of physics, the so-called `unified physical theory', if exists, would also correlate with set theory. For example, if sets are understood as any definable collections of objects, it looks quite suitable to consider the physical systems as sets of physical objects, afterwards one may consider the sets of the physical systems and so on up to the general set including all physical objects and systems, which would be the most general physical system of the unified physical theory. 

However, such ingenuous efforts to apply the set theory to physical description meet set-theoretical paradoxes similar to ones met by G.Cantor in his `naive' set theory \cite{Can874}. Indeed, already the set of sets of physical objects meets the Russell's antinomy (see, for example, \cite{Irv14}), because it is exactly the `set of all sets that are not members of themselves' and it appears to be a member of itself if and only if it is not a member of itself. Thus, a paradox. Moreover, the `all including set' corresponding to `unified physical theory' meets the Cantor's paradox, according to which, this set would be larger than itself, because by definition needs to contain its own power set (the set of all subsets) with higher cardinality. Thus, both a structure and an existence of the `unified physical theory' are under question. 

To avoid paradoxes, the numerous axiom systems were proposed for set theories, but it is not clear how they correlate with the physical objects and theories and whether it is possible to apply them to physical description at all. 

Physical description has some specifics, because, unlike mathematics, physics appeals to objective reality as a criterion of truth. Thus, to avoid logical and set-theoretical paradoxes in physical description, we will start from the detailed analysis of the process of cognition of objective reality in physics. 

\section{\bf Schemes of Cognition}
\label{sec:SC}

A cognition implies a separation on who is cognizing and what is being cognized. In physics it is an \emph{observer} who is cognizing the \emph{physical surroundings}, which is an \emph{objective reality} for him: 
\begin{description} 
\item[Postulate~1] The \emph{physical surroundings} is an objective reality for an \emph{observer} being in cognition of it. 
\end{description}

Using available methods of cognition $\widehat{P}$: $\{ \widehat{P}_k\}$ an observer identifies a \emph{physical object} $\frak{a}$ from the physical surroundings $\frak{R}$ as a collection of distinctive properties $\{p_l\}=\{p\} = \widehat{P} \frak{a}$, which are called a \emph{preset of predicates} and are exactly a \emph{model} $a$ of $\frak{a}$ or an \emph{object} $a$ corresponding to a physical object $\frak{a}$: 
\begin{equation} 
a = \{p\} = \widehat{P} \frak{a} \rightleftharpoons  \frak{a} \in \frak{R}. 
\label{a} 
\end{equation} 

\begin{description} 
\item[Definition~1] An \emph{object} is a collection of distinctive properties called a \emph{preset of predicates}, which is a \emph{model} of a \emph{physical object} identified by an observer in physical surroundings by means of available for him methods of cognition. 
\end{description} 

If some predicates of a model $A = \{p\}$ of a physical object $\frak{A} \in \frak{R}$ are interpreted by an observer as models $a_i \rightleftharpoons \frak{a}_i \in \frak{R}$, so $A$ is called a \emph{system} and one has for it: 
\begin{equation} 
A = \{p\} = \left\{ \bigcup_i a_i, f^A, r^A \right\} \rightleftharpoons \frak{A} \in \frak{R},   
\label{A} 
\end{equation} 
where the predicates $\{f^A\}$ are the rules of collecting of objects $a_i$ included in the system $A$ and $r^A$ are some other predicates. The rules of collecting $\{f^A\}$ unify objects $a_i$ in the system $A$ or interconnect them, so one may define $F^A = \{f^A\}$ as \emph{fields}. To be interconnected, the objects need to have the predicates \emph{conjugate} with the fields. 

\begin{description} 
\item[Definition~2.1] A \emph{system} is an object, in which some predicates are identified as other objects called \emph{included} into this system \emph{containing} them. 
\item[Definition~2.2] A \emph{field} of a system is the rules of collecting of objects \emph{conjugate} with this field and included into this system. 
\end{description} 

For example, the masses of the massive particles are predicates conjugate with gravitational fields interconnecting particles into the gravitational system. The electric charges are conjugate with electromagnetic fields, the massive charged particles are conjugate with both of these fields simultaneously. 

Objects $a_i$ in Eq.(\ref{A}) may be considered as systems $A_i$ containing one or more elements, so one can generalize Eq.(\ref{A}) for a system containing other systems: 
\begin{equation} 
A = \left\{ \bigcup_i A_i, f^A, r^A \right\} \rightleftharpoons \frak{A} \in \frak{R}, 
\label{AS} 
\end{equation} 
what represents a \emph{process of cognition} based on separation of objects from the physical surroundings and their integration into systems by means of some rules $\{f^A\}$. 
\begin{description} 
\item[Definition~3] A \emph{process of cognition} is a \emph{main} scheme of cognition based on a separation of objects from physical surroundings and on their integration into systems by means of the rules of collecting, what is a \emph{systematization} of representations. 
\end{description} 

The system $A$ in Eq.(\ref{AS}) contains the systems and may be included in other systems, which, in their turn, may contain and may be included in other systems and so on up to infinity. Thus, we will call the process of cognition \emph{open} at the top and at the bottom, i.e. in macro-scale of containing systems and in micro-scale of included systems. 

The rules of collecting in the containing and included systems may be different from each other, so it is hard to use the main scheme of cognition for a large number of physical objects and systems, what limits the possibilities of cognition and prediction of events. 

The process of cognition may be unified by introducing the \emph{elementary objects} and the \emph{fundamental fields}: 
\begin{description} 
\item[Definition~4] The main scheme of cognition is called \emph{basic} if any system in it may be represented by combinations of \emph{elementary objects}, which other objects cannot be identified in, \emph{conjugate} with \emph{fundamental fields}, which cannot consist of other fields. 
\end{description} 

Since the \emph{elementary objects} $\mathbf{p}_n$ and the \emph{fundamental fields} $\mathbf{f}_m$ can represent any system, so all predicates relate to elementary objects and fundamental fields, and the specific predicates $r^A$ in Eq.(\ref{AS}) are predicates conjugate with fields $\mathbf{f}_l$ external to a system $A$:  
\begin{equation} 
A = \left\{ \bigcup_{n,m} \Big\{ \mathbf{p}_n, \mathbf{f}_m \Big\}, \mathbf{f}^A_k, r^A({\mathbf{f}_l}) \right\} \rightleftharpoons \frak{A} \in \frak{R}.      
\label{AE} 
\end{equation} 
There is no limitation for containing systems, but included ones are limited by elementary objects, so a basic scheme of cognition is open at the top and \emph{close} at the bottom, even if a number of elementary objects and fundamental fields is infinite. 

The scheme of cognition will be close at the top, if there is introduced a \emph{general} system, which, if exists, contains all identified objects and, therefore, cannot be included in other systems. According to Eq.(\ref{AS}), one has for the general system $U$:
\begin{equation} 
U = \left\{ \bigcup_k A_k, \mathbf{C} \right\} \rightleftharpoons \frak{U} \in \frak{R},      
\label{U} 
\end{equation} 
where $\mathbf{C} = \{f^U\}$ is a field of the general system, and specific predicates $r^U$ conjugate with external fields are absent, because there are no fields external to the general system by definition. 
\begin{description} 
\item[Definition~5.1] A main scheme of cognition is called \emph{general} if a \emph{general} system containing all objects and systems identified by an observer is introduced in it.
\item[Definition~5.2] A field of a general system is called a \emph{pre-space}. 
\end{description} 
A general system cannot be included in other system, so it is close at the top and open at the bottom. 

The scheme of cognition will be close both at the top and at the bottom, if the general system, elementary objects and fundamental fields are introduced in it simultaneously. 
\begin{description} 
\item[Definition~6] The main scheme of cognition is called \emph{unified} if elementary objects, fundamental fields and a general system also called \emph{unified} are introduced in it. 
\end{description} 

All systems in the unified scheme of cognition, including the unified one, are completely determined on the \emph{basis} $[\mathbf{ p}_n, \mathbf{ f}_m, \mathbf{C}]$: 
\begin{equation} 
U = \left\{ \bigcup_{n,m} \Big\{\mathbf{p}_n, \mathbf{ f}_m \Big\}, \mathbf{C} \right\} \rightleftharpoons \frak{U} \in \frak{R}.      
\label{UB} 
\end{equation} 

Thus, there are introduced four schemes of cognition in physics: 
\begin{enumerate}
\item the main scheme open at the top and bottom; 
\item the basic scheme close at the bottom; 
\item the general scheme close at the top; 
\item the unified scheme close at the top and bottom. 
\end{enumerate}
These schemes are based on a process of cognition and, it will be shown later that all of them are used in physical researches.

\section{\bf Consistency}
\label{sec:C}

Unlike mathematics, an observer in physics has a criterion of truth: in accordance with Postulate~1, it is a reality, which is objective for him, so one may define:
\begin{description} 
\item[Definition~7] The representations and presets are \emph{consistent} and the predicates are \emph{noncontradictory}, if they correspond to objective reality on practice (at an experiment). 
\end{description} 

A definition of presets looks not so general than sets, because sets are understood as definable collections of any objects, but presets are collections only of models of physical objects, i.e. only of predicates determined by an observer. However, everything an observer may identify in physical surroundings are physical objects, but he exactly deals with his representations about these objects. Moreover, presets include only determined properties of physical objects, so exclude the indefinite ones out of consideration. Exactly this `more general' definition of sets is a reason of uncertainties and a source of paradoxes in set theories. 
\begin{description} 
\item[Assertion~1] The presets are the most general units in cognition, and, unlike sets, are free of paradoxes. 
\end{description}
Indeed, being predicates of predicates, the presets are always the `members of themselves', so the Russell's antinomy for the `set of sets that are not members of themselves' simply does not relate to presets. The Cantor's paradox for the `all including set', which by definition has to include its own power set, so has to have the cardinality higher than itself, is also eliminated for the `all including' preset, because it already contains the predicates of all `physically existing' interconnections between objects (see Eqs.(\ref{A},\ref{U})) and includes all `physically existing' subsystems, so coincides with its own `physically existing' `power preset'. 

Thus, presets are free of paradoxes and a general system may be consistently introduced with them. But how complete are such representations? May a general system contain all objects from physical surrounding, i.e. may it be a model of the physical surroundings: $U \rightleftharpoons \frak{R}$? 

According to Definition~2.1 and Eq.(\ref{U}), the general system corresponds to the physical object: 
\begin{equation} 
U \rightleftharpoons \frak{U} \in \frak{R},  
\label{UR} 
\end{equation} 
so if $U \rightleftharpoons \frak{R}$ then $\frak{U} \equiv \frak{R}$, i.e. a physical surroundings is a physical object. But a physical object by Definition~1 is identified as something separated from surroundings, so if a physical surroundings is a physical object, what, in this case, is it separated from? Only a part may be separated from a whole, so:
\begin{description} 
\item[Assertion~2] Any general system based on the process of cognition can represent only a part of physical surroundings. 
\end{description}

Thus, a general system may be consistently introduced by presets, but it may correspond only to some part of physical surroundings. It is fundamental limitation following from a process of cognition based on separation.

\section{\bf Logic of Cognition}
\label{sec:LC}

Using an objective reality as a criterion of consistency, one may also introduce a \emph{logic} of cognition: 
\begin{description} 
\item[Definition~8.1] A \emph{logic} is the generalized methods of cognition, which lead to consistent representations. 
\item[Definition~8.2] A \emph{logical system} is a general system, which a logic is introduced in.
\end{description} 

Being a general system, a logical system always represents only a part of physical surroundings, so one may assert:
\begin{description} 
\item[Assertion~3] A logical system is \emph{isolated} in frames of logic introduced in it. 
\end{description} 
Indeed, if an object $a$ is not an element of a logical system $U$, but logically interconnected with $U$, so $U$ is not general and is not logical. Thus, a contradiction. If $a$ is an element of $U$, but its interconnections with $U$ differ from logical ones, so $U$ is inconsistent and not logical. Thus, again a contradiction and an assertion is proved. 

It immediately follows from an isolation of a logical system: 
\begin{description} 
\item[Consequence~3.1] A logical system is \emph{self-consistent}: an observer, objects, fundamental fields and a pre-space in it are consistently interconnected only with each other.  
\item[Consequence~3.2] A logical system is \emph{self-defining}: `general' predicates of containing systems follow from `partial' ones of included systems defined in frames of general ones.
\end{description} 

A separation in a logical system is the getting partial from general, while an integration is the getting general from partial. In mathematics it is called a \emph{deduction} and \emph{induction} correspondingly. 

An introduction of elementary objects and fundamental fields in basic scheme is, in fact, a generalization of a process of cognition, but in basic scheme the general system is not introduced. In general scheme the rules of separation and integration are not generalized, so only a general system of unified scheme is a logical system: 
\begin{description} 
\item[Assertion~4] A logical system can be created only by means of the unified scheme of cognition. 
\end{description} 

\begin{figure*}[t]
	\begin{center}
		\includegraphics*[angle=0,width=155mm]{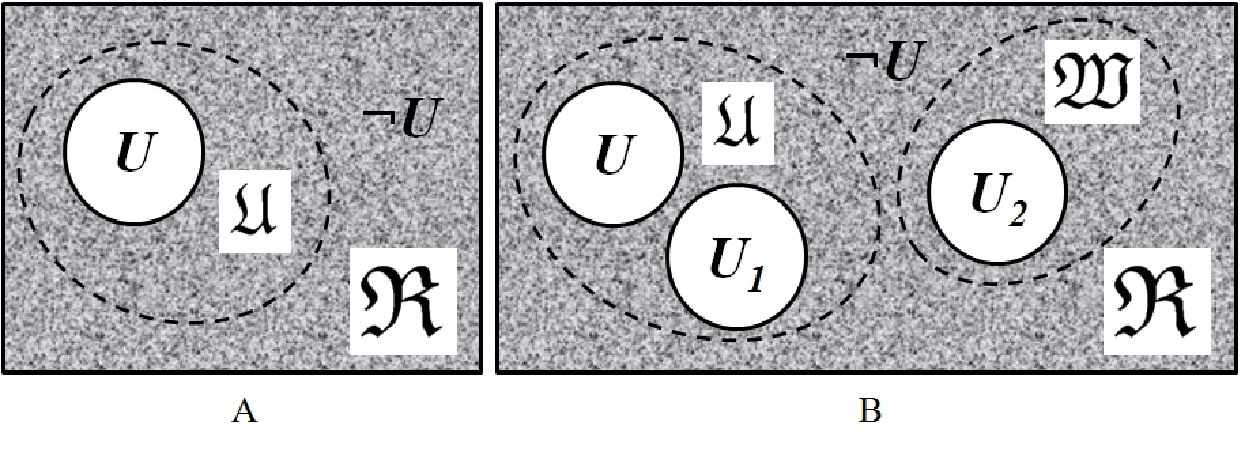}
	\caption{Binary (A) and Multiple (B) Logical Systems in Cognition}
	\label{fig:L}
	\end{center}
\end{figure*}

Since a logical system $U$ is logically isolated, its logical negation $\neg U$ may be introduced. The region $\neg U$ corresponds to an incognizable for $U$ part of physical surroundings, so a disjunction of $U$ and $\neg U$ corresponds to the physical surroundings (see Figure~\ref{fig:L}A): 
\begin{equation} 
\left( U \bigcup \neg U \right) \rightleftharpoons \frak{R}.  
\label{UR} 
\end{equation} 

Everything in physical surroundings has to be interconnected, because the `unconnected', if exists, might not influence an observer and could not be considered a part of his physical surroundings $\frak{R}$. For this reason, an existence of other `physical surroundings' $\frak{R}_1$, $\frak{R}_2$, etc., unconnected with $\frak{R}$, is only a subject for groundless speculations: 
\begin{description} 
\item[Postulate~2] There is only one objective reality in physics, which is physical surroundings. 
\end{description} 

The situation with the region $\neg U$ is quite different. From one side, a logical system $U$ is logically isolated from $\neg U$, but from other side, being the parts of the physical surroundings, regions $U$ and $\neg U$ need to be interconnected with each other. It means that the interconnections between $U$ and $\neg U$ exist, but exceed the logical ones introduced inside $U$. 

If interconnections different from logical ones exist, logical systems with different logic may also exist. Such different logical systems may correspond, for example, to the same physical object $\frak{U}$: $U_1 \rightleftharpoons \frak{U} \in \frak{R}$ or to different object $\frak{W}$: $U_2 \rightleftharpoons \frak{W} \in \frak{R}$ (see Figure~\ref{fig:L}B). Different logical systems were first introduced in \cite{NB07} as \emph{off-site continuums}. 

There is no reason to limit an amount of physical objects and logical systems, so:  
\begin{description} 
\item[Postulate~3] An infinite amount of different logical systems (i.e. general systems with different logic) may correspond to physical surroundings. 
\end{description} 

It does not mean that the logical way of cognition is useless. For example, though the cardinality of real numbers ($\aleph_1$) is higher than of rational ones ($\aleph_0$) and one cannot express irrationals by rationals, one can approximate any real number $x$ with any accuracy $\epsilon$ by corresponding rational number $q$: 
\begin{equation} 
\forall x \in \mathbb{R}, \forall \epsilon, \exists q \in \mathbb{Q}: | x-q | < \epsilon.
\label{QR} 
\end{equation} 

Thus, even if a logical system is logically isolated and physical surroundings may be corresponded to many logical systems, a logical way of cognition may be quite accurate in predictions and useful in cognition. 

\section{\bf Relativistic Theories}
\label{sec:RT}

In Euclidean geometry ``that which has no part'' is a \emph{point}, which in mathematics refers usually to an element of some collection, called space, generally more or less related to geometry. 

Being inside this paradigm, classical relativistic theories: Newtonian physics, special and general relativities understand a spacetime as a collection of points. All points in a spacetime have spacetime coordinates. 

Physical systems in relativistic theories are represented by elementary objects and fundamental gravitational and electromagnetic fields, and all of them are considered in a spacetime. Elementary objects in a spacetime are also points, because, otherwise, they would `have parts' and not be elementary. Elementary objects have \emph{mass} and \emph{charge} and are sources of fundamental fields. 

Thus, relativistic theories are systems of representations based on elementary objects, fundamental fields and a spacetime, so are exactly in frames of a unified scheme of cognition Eq.(\ref{UB}): 
\begin{description} 
\item[Assertion~5] Newtonian physics, special and general relativities are logical systems corresponding to the unified scheme of cognition.
\end{description}

An existence of the spacetime coordinates means an \emph{arrangement} of points, which predefines a consequence of events, so predefines the \emph{cause-effect chains} and a \emph{causality} in relativistic theories. To predict events, an arrangement needs to be continuous, so space and time in Newtonian physics, a spacetime in special relativity and a spacetime manifold in the general relativity have continuous axes of coordinates. 

Theories are called \emph{relativistic}, because a whole class inertial frames of references $\{S\}$: $S', S'',...$ interconnected with each other by Lorentz transformations are equivalent for physical description. Generally speaking, there may exist an infinite amount of classes of frames $\{\widetilde{S}\}$: $\widetilde{S}', \widetilde {S}'',...$ also interconnected with each other by Lorentz transformations, but only one of such classes $\{S\}$ has physical meaning. It is a fundamental problem of so-called `origin of inertiality', which in relativistic theories is unknown. 

The detailed analysis of relativistic theories from the point of view of introduced schemes of cognition gives clear answer to this fundamental question. Indeed, unlike a spacetime, a pre-space $\mathbf{C}$ of the unified system Eq.(\ref{UB}) is not an arranged collection of points, but their interconnections (Definition~2.2), i.e. an arrangement itself or a causality. Thus, inertial frames of references are equivalent, because they have the same causality determined by an interval $\rmd s'^2 = \rmd s''^2 = ... = \rmd s^2 $ and correspond to the same pre-space $\mathbf{C}$, so to the same causality in the general system $U (\rmd s) = U$, but considered in different spacetime variables: 
\begin{equation} 
U(\rmd s) = U'(\rmd s') = ... \equiv U(\mathbf{C}) \rightleftharpoons \frak{U} \in \frak{R}.  
\label{USR} 
\end{equation} 
Since the relativistic theories are logical systems (Assertion~5), they are logically isolated, so objects, an observer and a causality, corresponing to class of frames $\{S\}$, are interconnected in one logical system (Assertion~3, Consequence~3.1). Other classes of frames $\{\widetilde{S}\}$ noninertial to $\{S\}$ correspond to different causalities, so are inconsistent with logical systems of relativistic theories. 

In some sense, Postulate~3 is an expansion of relativistic principle onto different logical systems with different causalities corresponding to different classes $\{\widetilde{S}\}$ of frames noninertial to $\{S\}$. Nevertheless, to be realized, any logical system needs to correspond to the physical object ($\frak{U}$ or $\frak{W}$ on Figure~\ref{fig:L}) from physical surroundings. Thus, one may assert:
\begin{description} 
\item[Assertion~6] Any logical system is available for physical description, if it corresponds to a physical object from physical surroundings.  
\end{description}

In general relativity an interval depends on the distribution of the gravitating masses over the spacetime: $\rmd s^2 = \rmd s^2(x^\mu)$, where $x^\mu$ is the spacetime coordinates, so the causality depends on the spacetime coordinates, and a spacetime is a manifold: 
\begin{equation} 
U [\rmd s(x^\mu)] \equiv U[\mathbf{C}(x^\mu)] \rightleftharpoons \frak{U} \in \frak{R}.  
\label{UGR} 
\end{equation} 

The logical system in general relativity is also isolated, so Eq.(\ref{UR}) is true for it: $\left( U \bigcup \neg U \right) \rightleftharpoons \frak{U}$. Physically, the region $\neg U$ in Newtonian physics and special relativity is beyond a spacetime, and in general relativity it is beyond a horizon of events, i.e. exactly beyond a causality determined by a pre-space $\mathbf{C}$. In fact, different worlds, considered in modern physics behind the horizon of events of the black holes, are different logical systems introduced in Postulate~3. Of course, such worlds, if exist, are not the only ones. 

A logical isolation of consistent general system leads to conservation of some integral characteristics of all included objects and their interconnections. In general relativity it is expressed by a zero variation of actions of matter $\mathcal{S}_m$ and of the gravitational field $\mathcal{S}_g$: 
\begin{equation} 
\delta (\mathcal{S}_m + \mathcal{S}_g ) = 0. 
\label{Smg}
\end{equation} 
For example, a variation of this equation by the metric tensor $g_{\mu\nu}$ leads to well-known Einstein's field equations. 

In modern physics a \emph{universe} is usually understood as `all of spacetime and everything that exists therein' or as `the totality of existence, including all matter and energy'. After an introduction the different logical systems these definitions look uncertain or even contradictory. Indeed, `the totality of existence' probably corresponds to the physical surroundings $\frak{R}$, while `all of spacetime' to the logical system $U$, which is not the same. 

In fact, a universe is understood as everything available for cognition, as a reality cognizable for observers. If inertial observers are taking into account by default, so a universe is identified with a logical system $U$. `Other universes' also considered in some physical theories are called a \emph{multiverse}. 

According to expansioned relativistic principle (Assertion~6), not only inertial observers need to be taking into account, so:
\begin{description} 
\item[Definition~9] A \emph{universe} is a reality cognizable for observers.  
\end{description}
Thus, a universe is an aggregate of logical systems $\left( \bigcup U_i \right)$, which form its \emph{multi-space structure}. Such understanding of a universe was proposed in \cite{NB08, NB12}. 

In some theories a universe is considered correlated with a human brain: as a brain-like universe, as an existence of the consciousness in the world, as a Brain World. In our approach, a universe looks like a brain because a cognizable reality is a model of reality built by an observer by means of available for him methods of cognition, which is determined by his brain. In fact, a universe is a reflection of the physical surroundings onto the human brain. 

A cardinality of a continuum in set theory is $\aleph_1$ and of its power set (the set of all subsets) is $2^{\aleph_1}$. The presets correspond to objective reality, so includes not all possible, but all `physically existing' subsets (see Section~\ref{sec:LC}), therefore, a cognitive capacity of relativistic theories is more than the cardinality $\aleph_1$ of a continuum and cannot exceed the cardinality of its power set $2^{\aleph_1}$. 

\section{\bf Quantum Theories and Cosmology}
\label{sec:QTC}

In relativistic theories any system may be represented by the combinations of the point-like elementary objects and fundamental gravitational and electromagnetic fields, but the physical description of objects of quantum physics are based on the wave functions, which are incompatible with such representations. Nevertheless, objects of quantum physics are considered in a spacetime corresponding to the pre-space of the relativistic theories. Thus, in quantum physics the general system is introduced, but elementary objects are not used, so it is a general scheme of cognition (Definition~5.1): 
\begin{description} 
\item[Assertion~7.1] Modern quantum theories (QM, QED, QCD, ...) correspond to the general scheme of cognition open in micro-scale.  
\end{description}
Therefore, according to Assertion~4, quantum theories are not logical systems, but transition schemes (Eq.(\ref{U})) for searching the new natural laws in micro-scale of included systems. 

Observations in macro-scale in modern cosmology also look incomprehensible for relativistic theories. It is, for example, the accelerated expansion of the universe, the velocity distributions in galaxies, the gravitational lensing by clusters, the cosmic structure formation, etc. Cosmologists are trying to `explain' such effects by introduction of the new forms of matter: dark matter, dark energy, etc. In fact, it is the efforts to stay on representations of relativistic theories, so to be in frames of the basic scheme of cognition. Indeed, even the efforts to find `particles' corresponding to new forms of matter, for example, to dark matter, are nothing else but an effort to consider the new forms of matter as systems of elementary objects and fundamental fields.   
\begin{description} 
\item[Assertion~7.2] Modern cosmology corresponds to the basic scheme of cognition open in macro-scale. 
\end{description}

Also as quantum theories, the cosmological models are transition schemes for searching of new natural laws, but in macro-scale of the containing systems. However, if, for example, `particles' of dark energy cannot be introduced, so modern cosmology cannot be `explained' in frames of basic scheme of cognition, and the main scheme needs to be used nevertheless. The question is: May cosmological objects be `unified' in one `physical theory' in the future? Or, more generally: 
\begin{itemize} 
\item Can relativistic, quantum theories and modern cosmology be unified in one logical system, i.e. in `unified physical theory'?
\end{itemize}

\begin{description} 
\item[Assertion~8] Objects of relativistic, quantum theories and of modern cosmology are from different logical systems and cannot be considered in one `unified physical theory'. 
\end{description}

Since a pre-space is not a collection of points, but an arrangement of objects, the elementary objects do not need to be point-like. In mathematics there are known the spaces quite different from collections of points. For example, the functional spaces such as the Banach or Hilbert spaces, where the norm, the distance and the main operations like in Euclidean space are introduced for functions of special kinds. Apart from the classical Euclidean spaces, examples of Hilbert spaces are spaces of square-integrable functions, spaces of sequences, Sobolev spaces of generalized functions, Hardy spaces of holomorphic functions, etc. 

Generally speaking, the physical object $\frak{a}\in \frak{R}$ may be represented in any space if some operators (i.e. methods of cognition) $\widehat{Q} =\{ \widehat{Q}_m \}$ are introduced in it: 
\begin{equation}
a(q) = \{q\}= \widehat{Q} \frak{a} \rightleftharpoons \frak{a} \in \frak{R}.
\label{aQ}
\end{equation} 

When the object $a=\{p\}=a(p)$ from Eq.(\ref{a}) determined by methods $\widehat{P}$ corresponds to the same physical object $\frak{a}$, so: 
\begin{equation} 
a(p)= \widehat{P} \frak{a} \rightleftharpoons \frak{a} \rightleftharpoons \widehat{Q} \frak{a} = a(q).   
\label{apq} 
\end{equation} 

Thus, it has to be some correspondence between $a(p)$ and $a(q)$, and one may introduce a transition matrix $\mathcal{T}_{lk}$ for discrete models $a(p)=\{p_k\}$ and $a(q)=\{q_l\}$ or a transition function $\mathcal{T}(p,q)$ for distributions $a(p)$ and $a(q)$, for which: 
\begin{equation} 
p_k = \mathcal{T}_{kl} q^l, \quad a(p) = \int_q \mathcal{T}(p,q) a(q) \rmd q.  
\label{Wpq}
\end{equation} 
Here $\mathcal{T}_{kl}$ and $\mathcal{T}(p,q)$ need to satisfy to the conditions: $p_k = \mathcal{T}_{kl} \mathcal{T}^{lk} p_k$ and $a(p)=\int_q \rmd q \mathcal{T}(p,q)\int_p \rmd p \mathcal{T}(q,p)a(p)$. 

It may be, for example, the Fourier transformations and the Fourier integrals ($\mathcal{T}(p,q)= e^{\rmi 2\pi pq}$, $\mathcal{T}(q,p)= e^{-\rmi 2\pi pq}$) used for description of objects in quantum mechanics. 

If three or more different logical systems correlated with each other correspond to the same physical object: 
\begin{equation} 
a(p) \rightleftharpoons a(q) \rightleftharpoons a(r) \rightleftharpoons ... ,
\label{pqr} 
\end{equation} 
it is compatible for the QCD representations (see \cite{NB12} for details). 

On practice, different logical systems both in macro-scale of containing systems and in micro-scale of included ones are observed from the point of view of relativistic theories, so from the spacetime. Formally, to consider the cosmological objects, one needs to add to Eq.(\ref{Smg}) the action $S_M$ corresponding to systems containing the logical system of relativistic theories: 
\begin{equation} 
\delta (\mathcal{S}_m + \mathcal{S}_g + \mathcal{S}_M ) = 0.  
\label{SMmg}
\end{equation} 

Containing macro-objects influence to whole spacetime, so cannot be separated and identified inside it. Such influence are usually interpreted as a vanishing of the Ricci tensor: $R_{\mu\nu}= \Lambda g_{\mu\nu}$, where $\Lambda$ is a cosmological constant responsible for the expansion of the `universe' `explained' by introducing the \emph{dark energy}. Generally speaking, an influence of containing objects may be different in different regions of the spacetime, so depends on the spacetime coordinates $x^\mu$. Thus a cosmological constant is only an approximation to: $\Lambda = \Lambda(x^\mu)$ and it is a main scheme of cognition. 

Some included objects in the main scheme may be different logical systems, so their pre-spaces and causalities do not need to be in frames of a spacetime of relativistic theories. Indeed, spacetime representations are not appropriate for objects of quantum physics and they may also be not appropriate for some galaxes observed as `included' in our spacetime. In micro-scale the matter inside objects of quantum physics seems `captured' by strong interactions. In macro-scale the matter inside galaxes seems `attracting' by the \emph{dark matter}. Thus, the dark matter and strong interactions correspond to observations of objects of included systems from the containing one in macro- and micro-scales, while on the contrary the dark energy corresponds to observations of objects of containing systems from the included ones, i.e. from our spacetime (see \cite{NB12} for more details). 

Thus, relativistic and quantum theories, observation of modern cosmology may be successfully interpreted in frames of introduced schemes of cognitions. 

\section*{\bf Conclusion}

To understand the reason of paradoxes in set theories and to try to avoid such paradoxes in physical description, the process of cognition in physics were analysed. The presets, which are specific sets of predicates corresponding to the physical objects identified by an observer during cognition were introduced. Unlike sets, the presets are free of logical or set-theoretical paradoxes and may be consistently used in physical description. 

Four schemes of cognition in physics were introduced and considered: the main scheme open at macro-scale of containing objects and at micro-scale of included ones; the basic one close at micro-scale and open at macro-scale; the general one open at micro-scale and close at at macro-scale; the unified one close at macro- and micro-scales. All of these schemes are used in physical description, but only the last one corresponds to consistent logical system, so to consistent physical theory.  

A general system may be consistently introduced by presets, but cannot describe the whole physical surroundings, what is a consequence of a process of cognition based on separation. A logical system is logically isolated and the physical surroundings may contain an infinite amount of logical theories. Nevertheless, a logical way of cognition may be quite accurate in predictions and useful in cognition. 

The objects of quantum physics and of modern cosmology correspond to different logical systems and cannot be described consistently in so-called `unified physical theory', but may be successfully interpreted in frames of introduced schemes of cognitions. 


\end{document}